\documentclass[prb,aps,twocolumn,amsmath,amssymb,floatfix,
superscriptaddress]{revtex4-2}
\usepackage[dvips]{graphics}
\usepackage[justification=raggedright]{caption}
\usepackage{subcaption}
\usepackage{color}
\usepackage{float}
\usepackage{natbib}
\setcitestyle{numbers}
\setcitestyle{square}
\definecolor{dred}{rgb}{0,0,0.6}
\usepackage{graphicx}    
\usepackage{bm}  
\usepackage{amssymb}  
\usepackage{amsmath}
\usepackage{braket}
\usepackage[mathscr]{euscript}
\usepackage[colorlinks=true, allcolors=blue]{hyperref}
\begin{document}
\title{Higher order topology in a Creutz ladder}

\author{Srijata Lahiri and Saurabh Basu \\ \textit{Department of Physics, Indian Institute of Technology Guwahati-Guwahati, 781039 Assam, India}}

\date{\today}
\begin{abstract}
A Creutz ladder, is a quasi one dimensional system hosting robust topological phases with localized edge modes protected by different symmetries such as inversion, chiral and particle-hole symmetries. Non-trivial topology is observed in a large region of the parameter space defined by the horizontal, diagonal and vertical hopping ampitudes and a transverse magnetic flux that threads through the ladder.  In this work, we investigate higher order topology in a two dimensional extrapolated version of the Creutz ladder. To explore the topological phases, we consider two different configurations, namely a torus (periodic boundary) and a ribbon (open boundary) to look for hints of gap closing phase transitions. We also associate suitable topological invariants to characterize the non-trivial sectors. Further, we find that the resultant phase diagram hosts two different topological phases, one where the higher order topological excitations are realized in the form of robust corner modes, along with (usual) first order excitations demonstrated via the presence of edge modes in a finite lattice, for the other.
\end{abstract}

\maketitle

\begin{center}\section{\label{sec:level1}Introduction}\end{center}
\par Topological insulators are materials that exhibit a gapped bulk resembling an insulator and an edge or a surface which is metallic \cite{Murakami_2011,RevModPhys.82.3045}. The conducting edge or the surface modes are robust and are protected by the symmetries of the system. They are oblivious to local perturbations and stay stable as long as the fundamental symmetries of the system are unaltered. Creutz ladder is one such quasi-1D system consisting of two legs of sites that are coupled by diagonal ($D$), horizontal ($L$) and vertical ($R$) hopping amplitudes \cite{PhysRevLett.83.2636,article,Gholizadeh_2018,Li2013}.\begin{figure}
    \begin{subfigure}[b]{\columnwidth}
         \includegraphics[height=34mm,width=73mm]{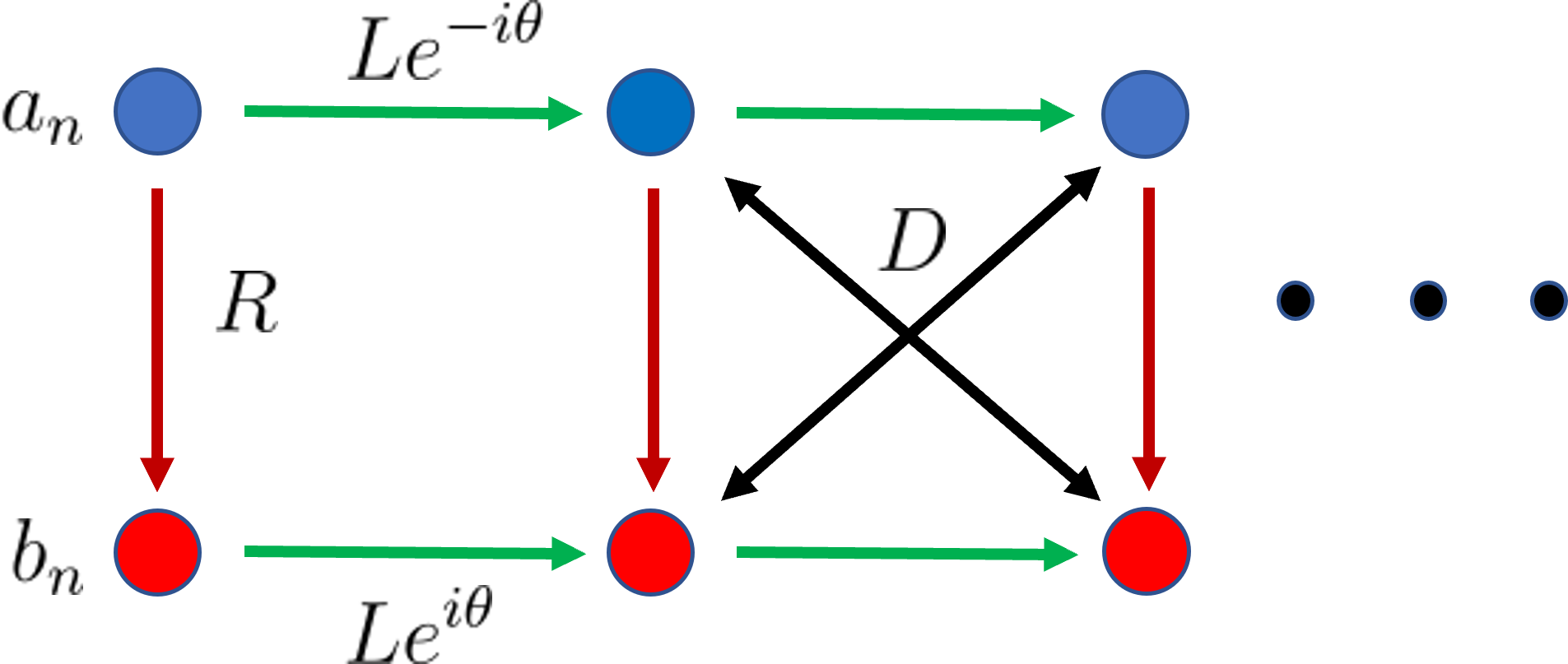}
         \caption{}
         \label{15.1}
     \end{subfigure}
      \begin{subfigure}[b]{\columnwidth}
         \includegraphics[height=75mm,width=\columnwidth]{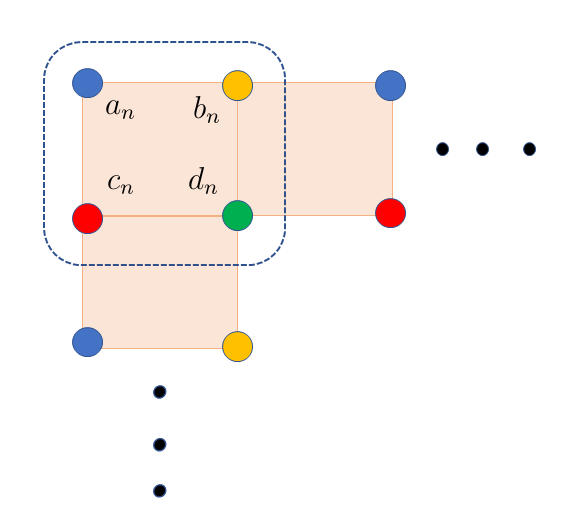}
         \caption{}
         \label{15.2}
     \end{subfigure}
 \caption{{(a) shows a schematic plot of the 1D Creutz ladder. $a_n$ and $b_n$ denote the two different sublattices. The different types of hopping amplitudes that characterize the ladder are also shown. (b) is a schematic representation of the 2D extrapolated version of the Creutz ladder. It has four sublattices namely $a$, $b$, $c$ and $d$ which are highlighted by different colours.}} 
 \label{fig15}
\end{figure}
 Additionally, a magnetic flux penetrates the ladder in a plane perpendicular to it. We use a Landau gauge to characterize the magnetic field. As a result, the horizontal hopping amplitudes carry a phase along with them (Fig. \ref{15.1}). This complex phase leads to destructive interference of the hopping amplitudes, and hence the model shows localization of particles corresponding to certain regions of the parameter space defined by $D$, $L$ and $R$. The topological character of this model is exhibited via the localized zero energy edge modes that remain confined to the two edges of the ladder (under open boundary conditions) in a certain parameter regime. This model being quasi-1D, it is difficult to place it into the conventional \textit{ten fold classification} of symmetries introduced by Altland and Zirnbauer \cite{PhysRevB.55.1142}. It is interesting to note that the zero modes of the Creutz ladder are results of both Aharonov Bohm caging and the topological character of the model. In the presence of this dual protection, the edge modes are robust even for a small system size. They find applications in the theory of quantum information due to their robustness.\par Higher order topology is a relatively new sub-field of topological insulators that is being actively explored in recent times \cite{doi:10.1126/sciadv.aat0346,PhysRevResearch.3.L042044,PhysRevLett.123.256402,PhysRevB.96.245115,PhysRevLett.119.246402,PhysRevB.97.155305,PhysRevLett.120.026801,Costa2021,Noguchi2021}. Higher order topological insulators (HOTI) refer to states of matter that show insulating behaviour both in the bulk as well as on the surface. In this case, topological non-triviality, in the form of robust gapless excitations arise in dimensions less than $d-1$, for a bulk that is $d$ dimensional. Therefore, HOTI exhibits the presence of corner modes in two dimensions and hinge modes in three dimensions as a signature of this non-trivial higher order topology. There are different approaches of arriving at the higher order excitations. The most general approach is to start with a conventional topological insulator (TI) and use a spatially dependent mass term to gap out the edge modes \cite{PhysRevB.100.115403}. The corner or the hinge of the system at which this mass changes sign binds the stable zero energy corner or hinge states which are resistant to minor perturbations as long as the relevant crystal symmetries are maintained \cite{PhysRevD.13.3398,PhysRevLett.106.106802}. Furthermore, the material candidates showing double band inverson properties are prospective systems that can show higher order topological behaviour \cite{PhysRevB.101.245110}. In this work, we shall exploit the chiral symmetry of the Creutz ladder which plays a crucial role in realising higher order corner modes in a two dimensional extrapolation of the model. We show a schematic diagram of our system in which different sublattices are shown via distinct colours (Fig. \ref{15.2}). The Creutz ladder possesses a chiral symmetry under the constraint of half a unit of magnetic flux penetrating each plaquette of the ladder (the unit of flux is taken as $\Phi_0=\frac{h}{e}$).\begin{figure}
\centering
\includegraphics[width=.75\columnwidth]{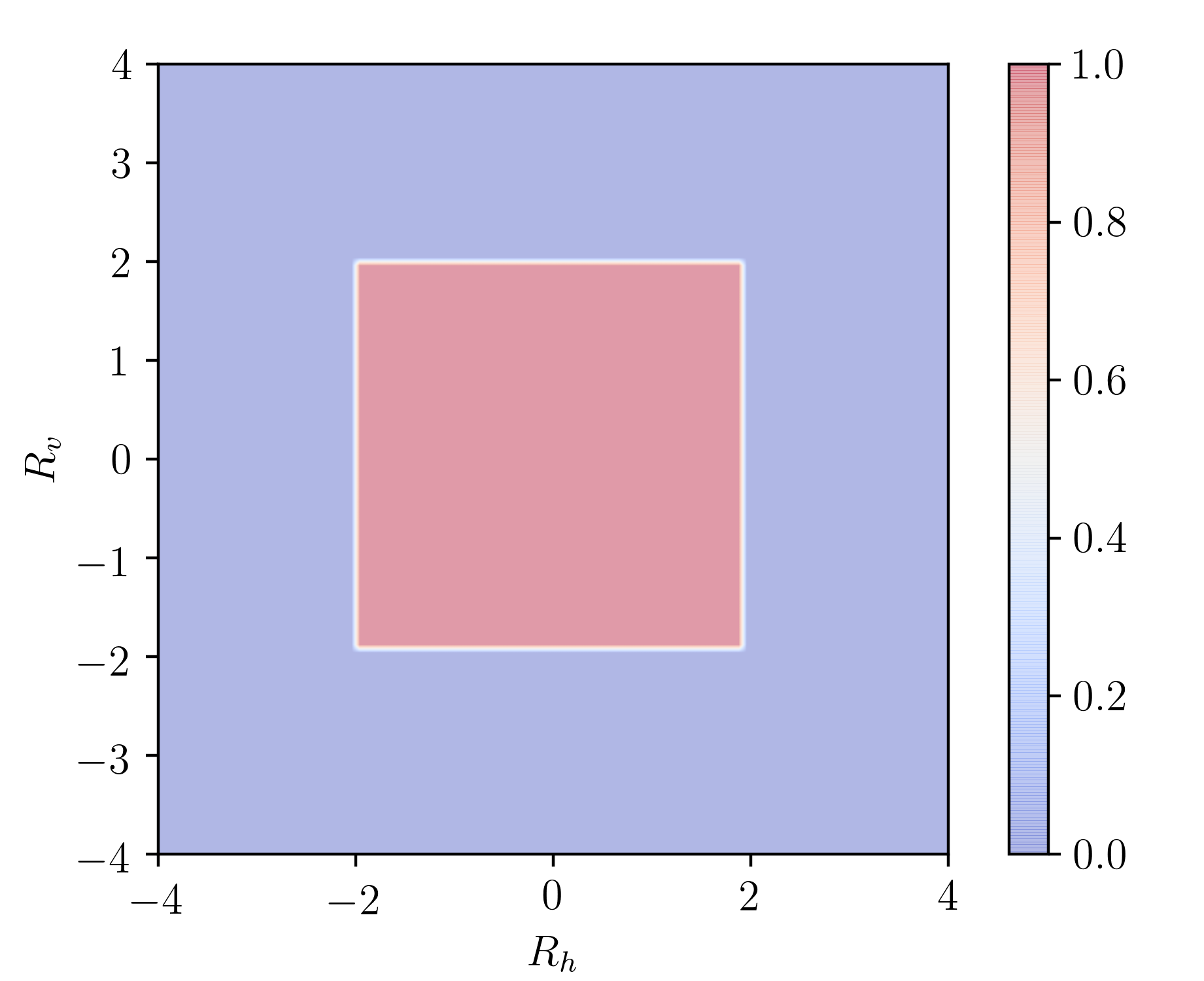}
 \caption{{The phase space plot shows the variation of winding number $\nu_{2D}$ with respect to the parameters $R_h$ and $R_v$. The winding number is non-trivial only when both $H_x(k_x)$ and $H_y(k_y)$ are in the topological phase, i.e $|R_h|,|R_v|< 2$.}}
\label{fig13}
\end{figure}
Maintaining this constraint, we build a two dimensional Hamiltonian and study its behaviour for different values of its parameters. The phase diagram shows two different topological regimes, one characterized by the presence of corner modes and the other by the edge modes. We associate two different topological invariants to characterize these phases.
\begin{figure}
    \begin{subfigure}[b]{\columnwidth}
         \includegraphics[height=65mm,width=\columnwidth]{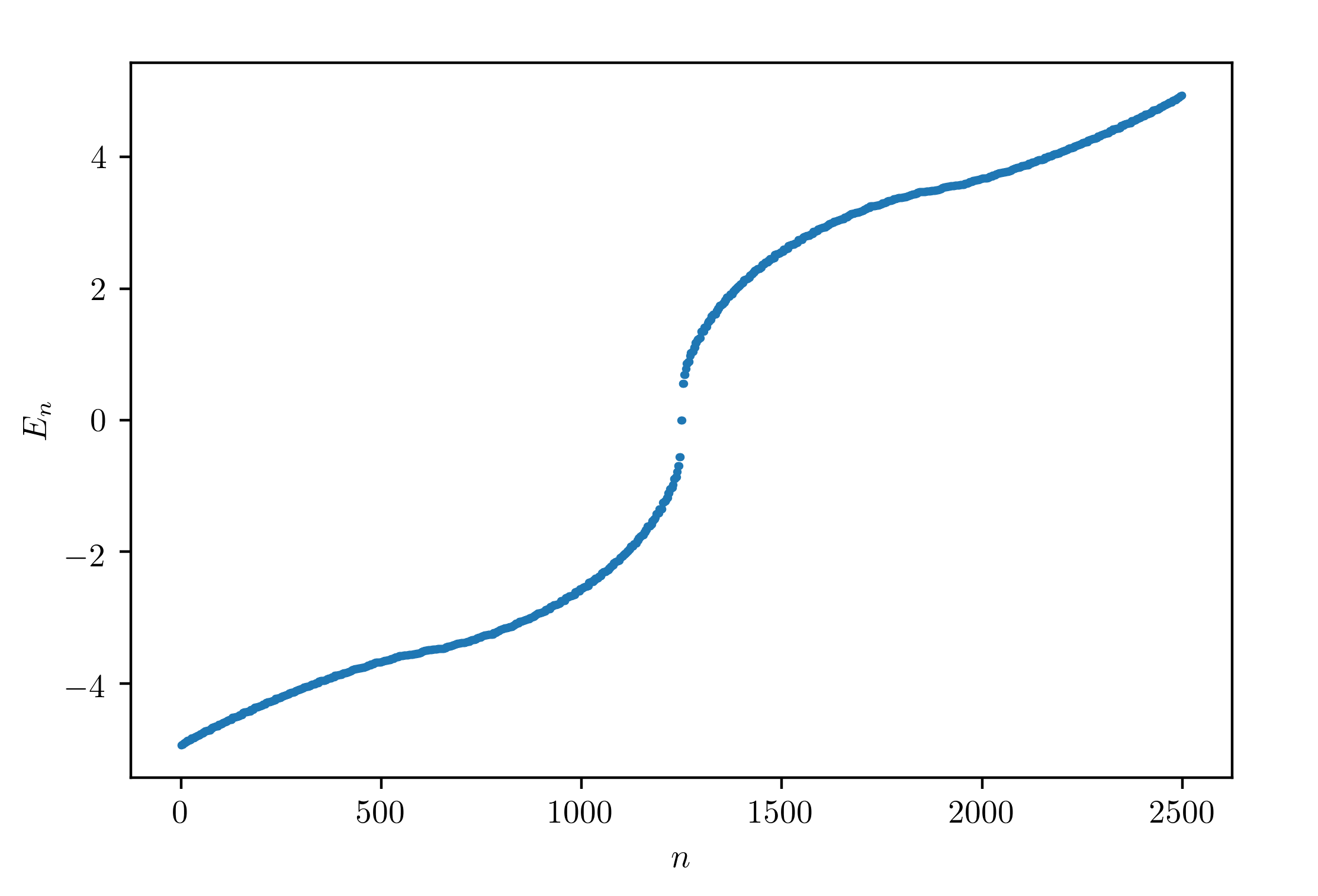}
         \label{2.1}
     \end{subfigure}
\caption{{This plot shows the energy eigenvalues for a $25\times25$ square lattice. The real space Hamiltonian is given in Eq. \ref{7}. The values of the parameters are $|R_h|=|R_v|=1.5$. Also $D_h=D_v=L_h=L_v=1$.}} 
\label{fig2}
\end{figure}
We also study the behaviour of the bulk energy spectrum along with the energy band structure of a ribbon-like configuration and analyse the various topological phase transition points.\par
The layout of the subsequent discussion is as follows. In section II we describe the model Hamiltonian and illustrate various key points that will aid us to arrive at a solution and understand the topological properties. In section III we discuss the results obtained by us. We elaborate on the phase diagrams and the topological invariants specific to these phases.
\begin{center}{\section{\label{sec:level2}The Hamiltonian}}\end{center}
Let us first fix the preliminaries of the Creutz ladder to facilitate subsequent discussions. As already mentioned, the Creutz ladder in its original form is a quasi-1D ladder system with horizontal, vertical and diagonal hoppings.\begin{figure}
          \begin{subfigure}[b]{\columnwidth}
         \centering
         \includegraphics[height=65mm,width=\columnwidth]{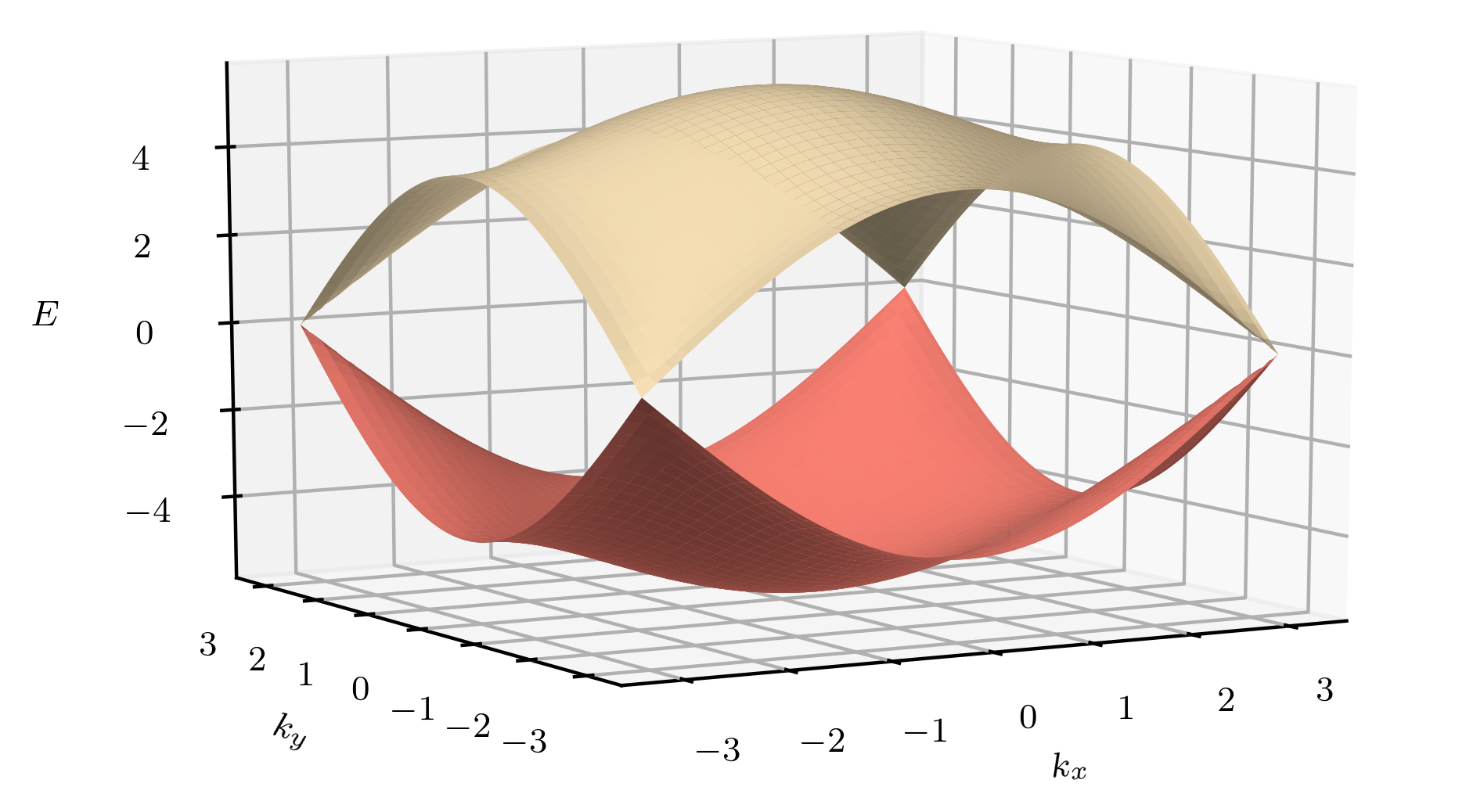}
         \caption{}
         \label{1.1}
     \end{subfigure}
     \begin{subfigure}[b]{\columnwidth}
         \includegraphics[height=65mm,width=\columnwidth]{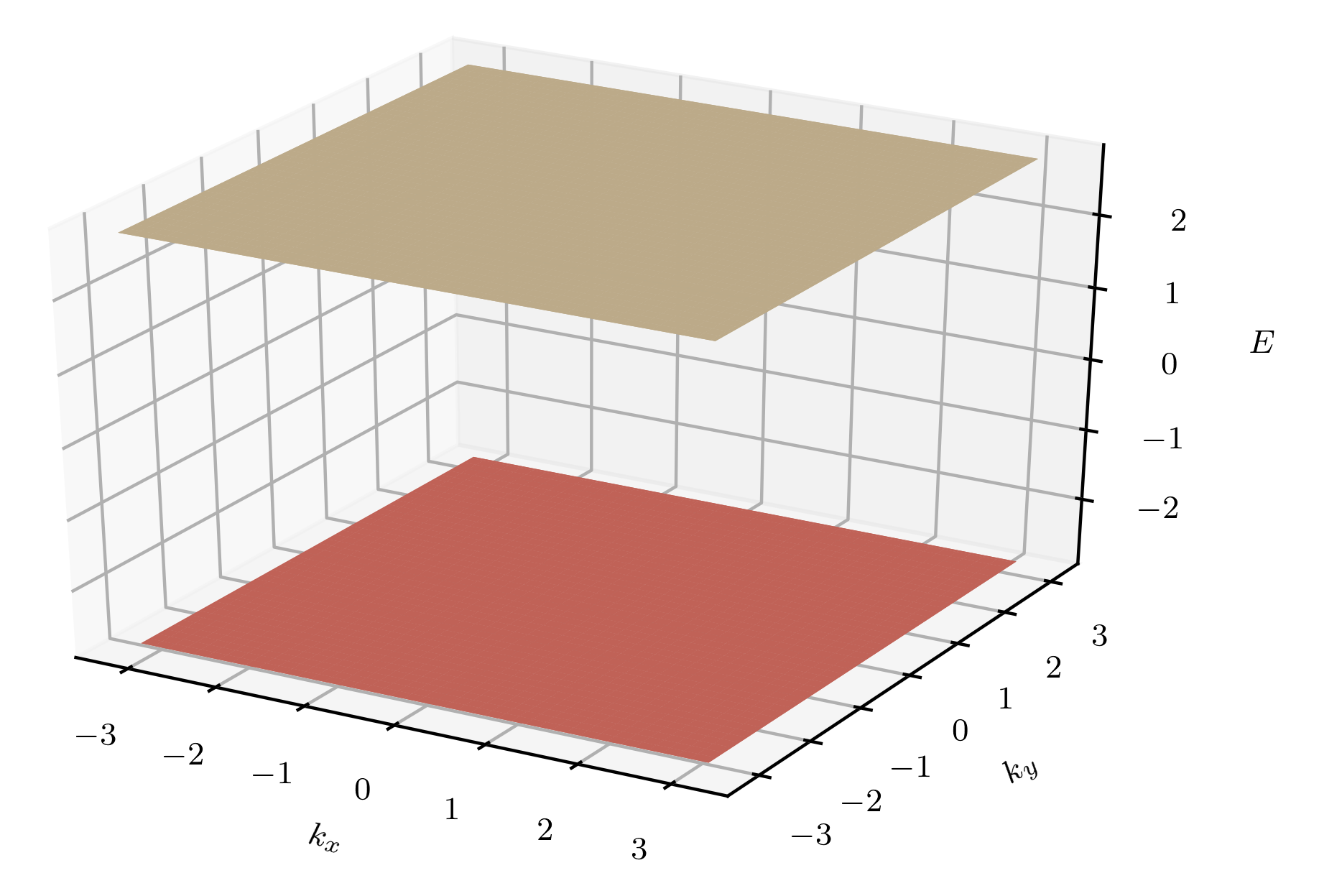}
         \caption{}
         \label{1.2}
     \end{subfigure}
\caption{{(a) refers to the energy spectrum of the bulk Hamiltonian Eq. \ref{4}. The bands are two fold degenerate. The values of the parameters are $|R_h|=|R_v|=2$, $D_h=D_v=L_h=L_v=1$. Gap closing occurs at the points $(k_x,k_y)$ = $(-\pi,-\pi)$, $(\pi,-\pi)$, $(-\pi,\pi)$, $(\pi,\pi)$. (b) depicts the flat band spectrum which occurs at $|R_h|=|R_v|=0$. The corner modes are most highly localized at the flat band points.}} 
 \label{fig1}
\end{figure}
\begin{figure}
\begin{subfigure}[b]{\columnwidth}
         \centering
         \includegraphics[height=65mm,width=\columnwidth]{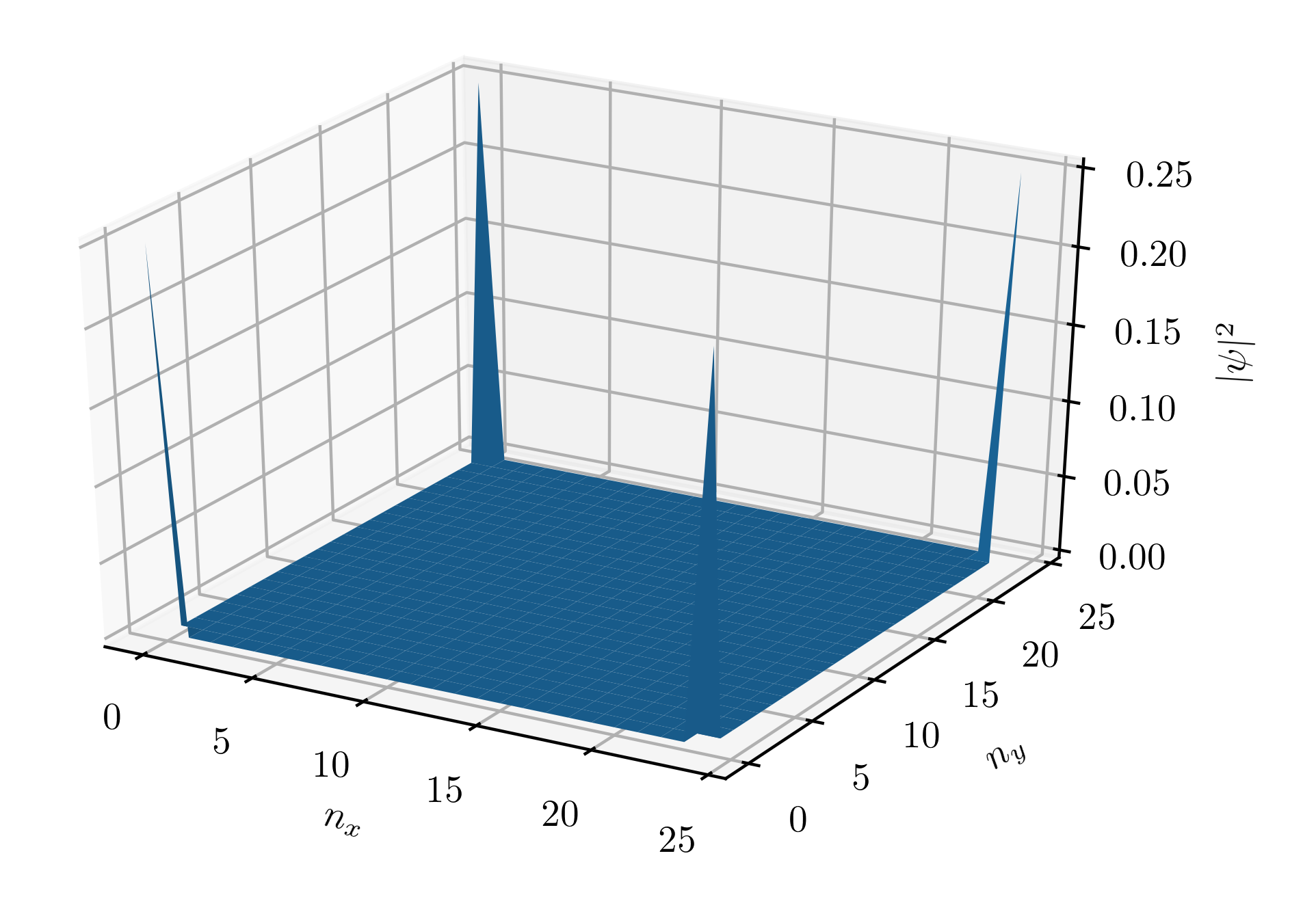}
         \caption{}
         \label{3.1}
     \end{subfigure}
\begin{subfigure}[b]{\columnwidth}
         \centering
         \includegraphics[height=65mm,width=\columnwidth]{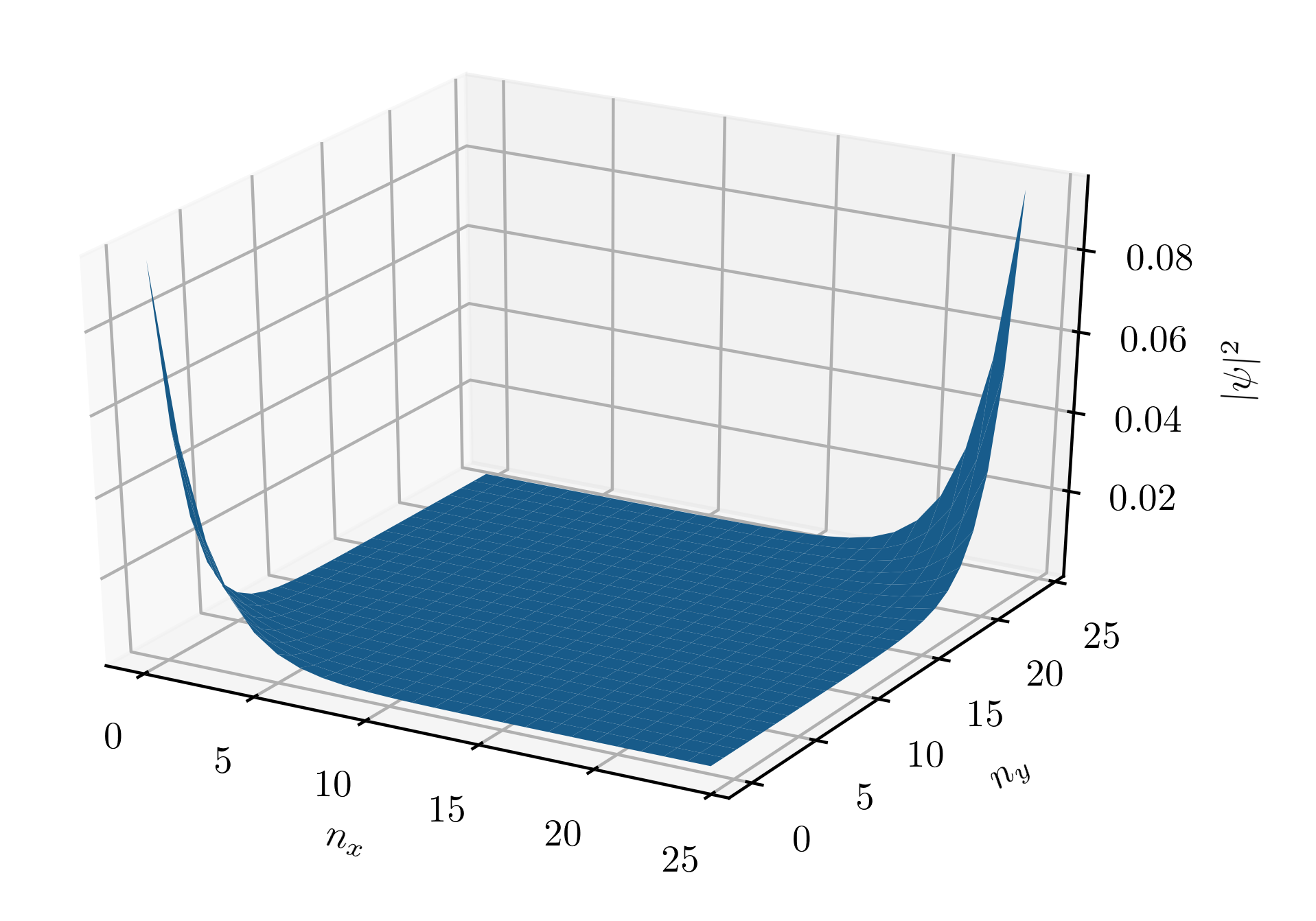}
         \caption{}
         \label{3.2}
     \end{subfigure}

\caption{{(a) shows the probability distribution for the zero energy corner modes for a $25\times25$ square lattice. The values of the parameters are $|R_h|=|R_v|=0$, $D_h=D_v=L_h=L_v=1$. It is seen that these modes are perfectly localized at the four corners of the lattice for the flat band point. (b) shows the same probability distribution for $|R_h|=|R_v|=1.5$, $D_h=D_v=L_h=L_v=1$.}} 
 \label{fig3}
\end{figure}
The magnetic field that penetrates the ladder perpendicular to the plane, gives us an extra degree of freedom in the horizontal hopping which picks up a Peierls phase. Further, there are two sublattices denoted as $a_n$ and $b_n$ within each unit cell of the ladder (here $n$ denotes the number of the unit cell which is also the $n^{th}$ rung of the ladder). The real space Hamiltonian of the original Creutz ladder is as follows,
\begin{align}
\begin{split}
H = &-\sum_n L(e^{i\theta}a_{n}^{\dagger}a_{n+1} + e^{-i\theta}a_{n+1}^{\dagger}a_{n} +e^{-i\theta}b_{n}^{\dagger}b_{n+1} \\&+e^{i\theta}b_{n+1}^{\dagger}b_{n}) +D(a_{n}^{\dagger}b_{n+1}+b_{n+1}^{\dagger}a_{n}+a_{n+1}^{\dagger}b_n\\&+b_{n}^{\dagger}a_{n+1})+R(a_n^{\dagger}b_n+b_n^{\dagger}a_n)
\end{split}
\end{align}
Here $L$ and $R$ denote hopping amplitudes along the leg and the rung of the ladder respectively. $D$ denotes hopping along the diagonal of a particular plaquette. $\theta$ is the Peierls phase introduced by the external magnetic field. If $\Phi$ denotes the total flux through each plaquette then, $ 2\theta=\frac{\Phi}{\Phi_0}$, where $\Phi_{0}$ denotes the magnetic flux quanta.\\
We fourier transform this Hamiltonian using the unitary transformation,
\begin{equation}
\label{2}
\begin{aligned}
    a_n&=\sum_k a_ke^{ikx_n}\\
    \end{aligned}
\end{equation}
Here $a_n$ ($a^{\dagger}_n$), represents the annihilation (creation) operators in the real space. The Hamiltonian in the momentum space now reads,
\begin{align}
\begin{split}
\label{3}
H(k)=2L\cos(k)\cos(\theta)\sigma_0&+2L\sin(k)\sin(\theta)\sigma_z\\&+(R+2D\cos(k))\sigma_x
\end{split}
\end{align}
The basis used in this case is $(a_k, b_k)$. Here $\sigma_{i=x,y,z}$ denotes the Pauli matrices. It is important at this point to understand the symmetries of the system \cite{PhysRevB.92.085118}. The model has an inherent inversion symmetry with respect to an axis that lies between the two legs of the ladder \cite{PhysRevB.83.245132}. It is expressed by the relation $\sigma_xH(k)\sigma_x=H(-k)$. Furthermore, it possesses a chiral symmetry that is illustrated via $\sigma_yH(k)\sigma_y=-H(k)$, which is true when $\theta=\frac{\pi}{2}$. The chiral symmetry is broken for other values of the phase $\theta$. As previously mentioned there is a time reversal symmetry inherent in the model which is maintained throughout the parameter space inspite of the presence of a magnetic field. It is given by $\sigma_xH^*(k)\sigma_x=H(-k)$. There is a debate in literature whether this, at all, should be called a time reversal symmetry \cite{PhysRevX.7.031057,PhysRevB.99.054302}. Lastly a particle hole symmetry exists in the system for $\theta=\frac{\pi}{2}$, which is expressed via, $\sigma_zH^*(k)\sigma_z=-H(-k)$. \par
For our purpose, the chiral symmetry of the model is of prime importance. Hence we stick to a particular value of the phase namely, $\theta=\frac{\pi}{2}$. The bulk Hamiltonian in the Fourier space is written as,
\begin{equation}
\label{4}
H(k_x,k_y) = H_x(k_x)\otimes\sigma^{y}_y + \mathbb{I}_x\otimes H_y(k_y)
\end{equation}
where, $H_x(k_x)$ and $H_y(k_y)$ are bulk Hamiltonians representing the Creutz ladder in the \textit{x} and \textit{y} directions respectively . The chiral symmetry operator for this model is given by $\sigma_T=\sigma^{y}_x\otimes\sigma^{y}_y$ \cite{PhysRevB.100.235302}. Again $\sigma^{y}_x$ and $\sigma^{y}_y$ represent chiral operators along the \textit{x} and \textit{y} directions respectively. Note that the suffix of the Pauli matrix $\sigma$ denotes the direction. It is obvious from the structure of the Hamiltonian (Eq. \ref{4}) that it anti-commutes with the chiral operator $\sigma_T$. This implies $\{H(k_x, k_y),\sigma_T\}=0$, thus conforming with the condition for chiral symmetry. \par The original Creutz ladder as mentioned in Eq. \ref{3} shows a gap closing transition for the set of values of parameters, namely $L=1,D=1,R=\pm 2$ and $\theta=\frac{\pi}{2}$. Gap closure is the signature of a topological phase transition. For $|R|< 2$, the Creutz ladder is in the topological phase and shows the presence of edge modes in the open boundary condition. The model belongs to the symmetry class BDI, which in one dimension is characterized by a $\mathbb{Z}$ type topological invariant. The invariant in this case (Eq. \ref{3}) is the winding number \cite{PhysRevB.78.195125,Ryu_2010}. If a given Hamiltonian can be written in the form,
\begin{equation}
    H(k)={d}_0\sigma_0 + \vec{d}(k)\cdot\vec{\sigma}
\end{equation}
where $\vec{d}(k)$ contains any two of the three components $d_x$, $d_y$, $d_z$, then a winding number can be defined for it. For our 1D model the winding number is given as,
\begin{equation}
\label{6}
    \nu=\frac{1}{2\pi}\int_{0}^{2\pi}\frac{d_{z}d(d_x)-d_xd(d_z)}{d_x^2+d_z^2}dk
\end{equation}
This winding number being a topological invariant remains unchanged unless the system goes through a gap-closing transition. For $|R|< 2$, the winding number $\nu = 1$ and the phase is topological. $\nu=0$ otherwise and indicates a trivial phase. \par 
It is important to understand why a gap closing transition is important for the formation of the corner or the edge states. The energy spectrum looks apparently similar on either side of the gap closing transition point. But the vital topological information is retained in the winding number ($\nu_{2D}$). Mathematically, the winding number number quantifies how many times a vector wounds around the origin as a function of a periodic parameter (as is evident from Eq. \ref{6}). It is a robust topological invariant in the sense that it is difficult to modify this number unless a drastic perturbation is applied to the system that changes the behaviour of the Hamiltonian completely. Physically, the winding number carries the same information as the Berry phase which is a phase picked up by a wave function as it is smoothly varied with respect to certain parameters \cite{Asb_th_2016,RevModPhys.82.1959,RevModPhys.66.899}.
\begin{align}
\begin{split}
    \gamma=\int_{{\alpha}_0}^{\alpha_t}\langle \psi(\alpha)|\nabla_\alpha|\psi(\alpha)\rangle d\alpha
\end{split}
\end{align}
Here $\alpha=\{\alpha_1,\alpha_2,...\}$ represents the parameters on which the wave function $\psi$ depends. In our case the crystal momentum $\vec k$ is the necessary parameter. A Hamiltonian with a non-trivial winding number cannot be adiabatically connected to an atomic insulator limit unless a gap-closing is involved. A small change in the Hamiltonian may change the shape of the path followed by the wave vector in the Brillouin zone, but it does not change the winding. Physically, in our model, this non-triviality is seen as the emergence of stable corner or edge modes.\par
\begin{figure}
\begin{subfigure}[b]{\columnwidth}
         \centering
         \includegraphics[height=65mm,width=\columnwidth]{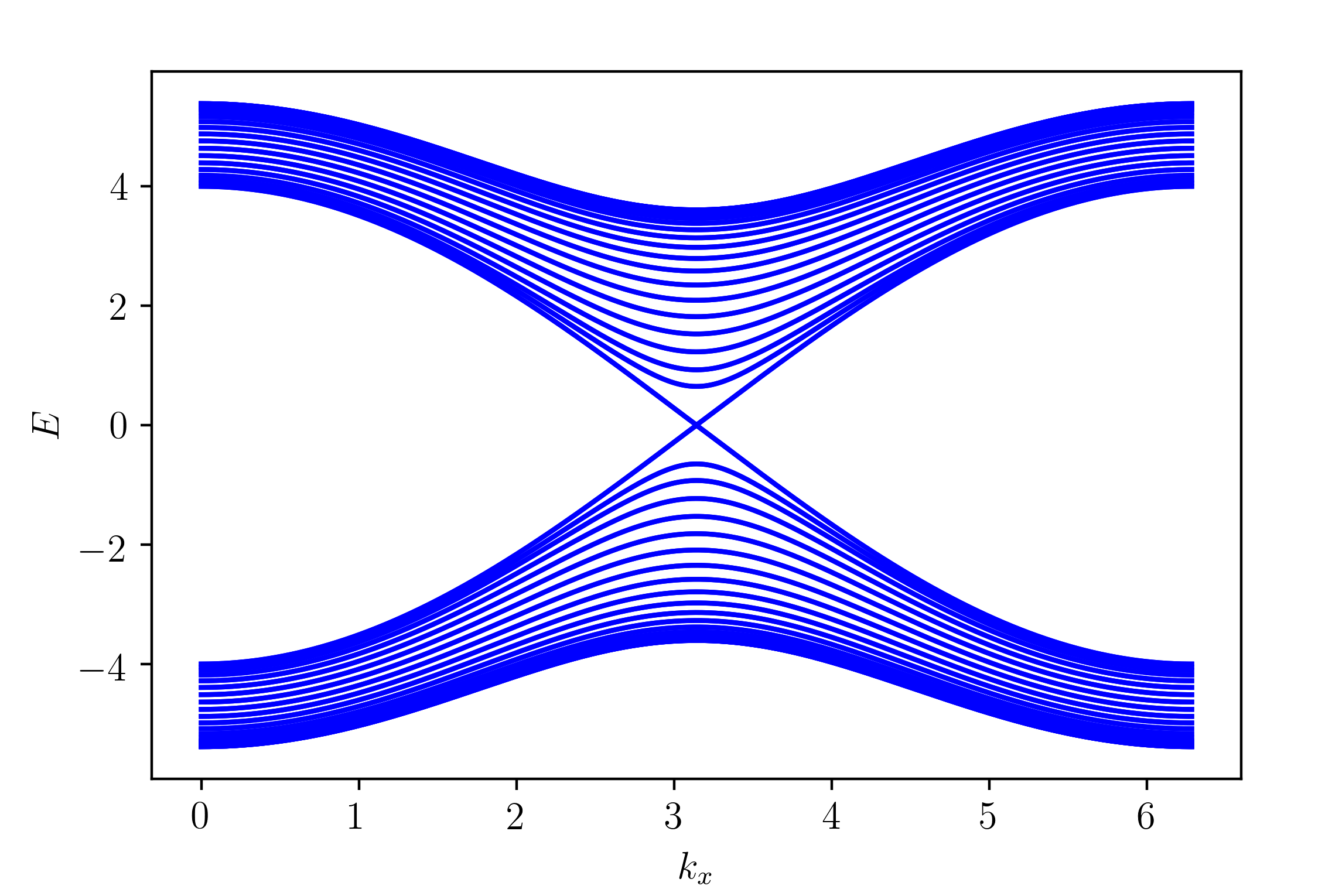}
         \caption{}
         \label{4.1}
     \end{subfigure}
\begin{subfigure}[b]{\columnwidth}
         \centering
         \includegraphics[height=65mm,width=\columnwidth]{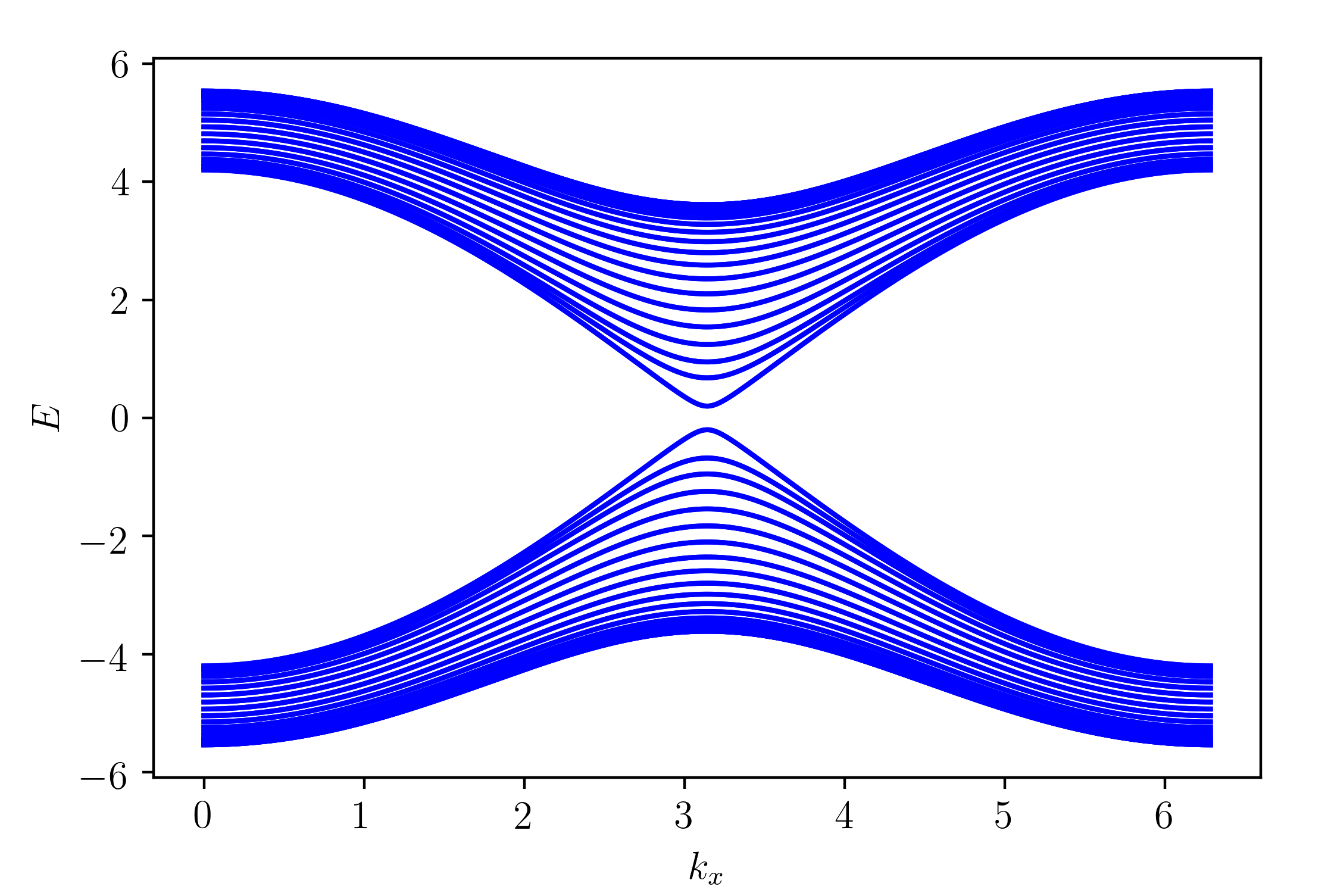}
         \caption{}
         \label{4.2}
     \end{subfigure}
\caption{{(a) denotes the bandstructure for the semi-infinite configuration (Eq \ref{9}). The Hamiltonian $H_y(k_y)$ is topological here. The values of the parameters are $|R_h|=2$, $|R_v|=1.5$. All the other parameters are unity. A gap closing transition is observed at this point. (b) shows the opening of the bulk gap as $R_h$ exceeds a value of $2$. Here the value of $R_h=2.2$. All other parameters are kept unchanged.}}
 \label{fig4}
\end{figure}
\begin{figure}
\begin{subfigure}[b]{\columnwidth}
         \centering
         \includegraphics[height=65mm,width=\columnwidth]{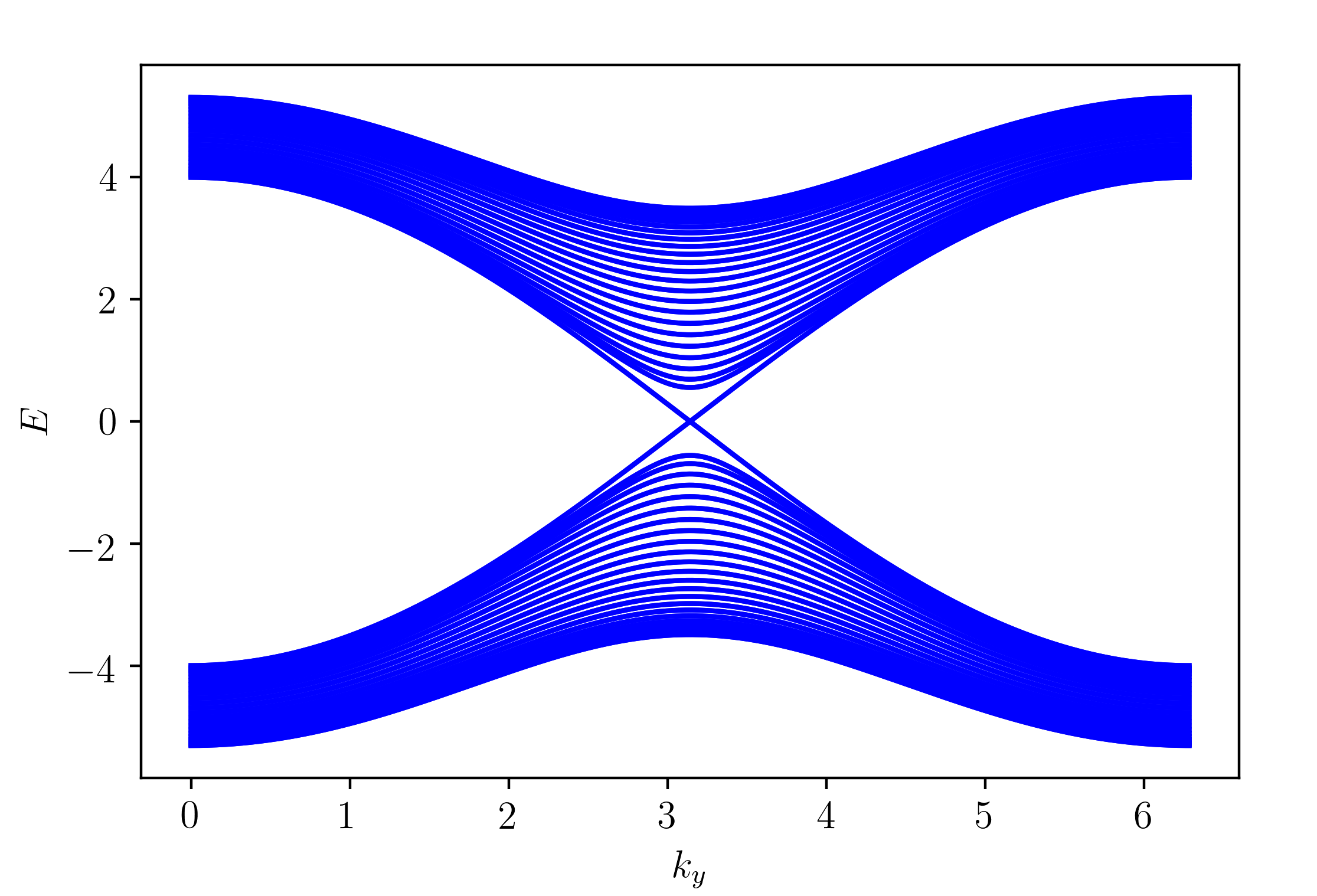}
         \caption{}
         \label{5.1}
     \end{subfigure}
\begin{subfigure}[b]{\columnwidth}
         \centering
         \includegraphics[height=65mm,width=\columnwidth]{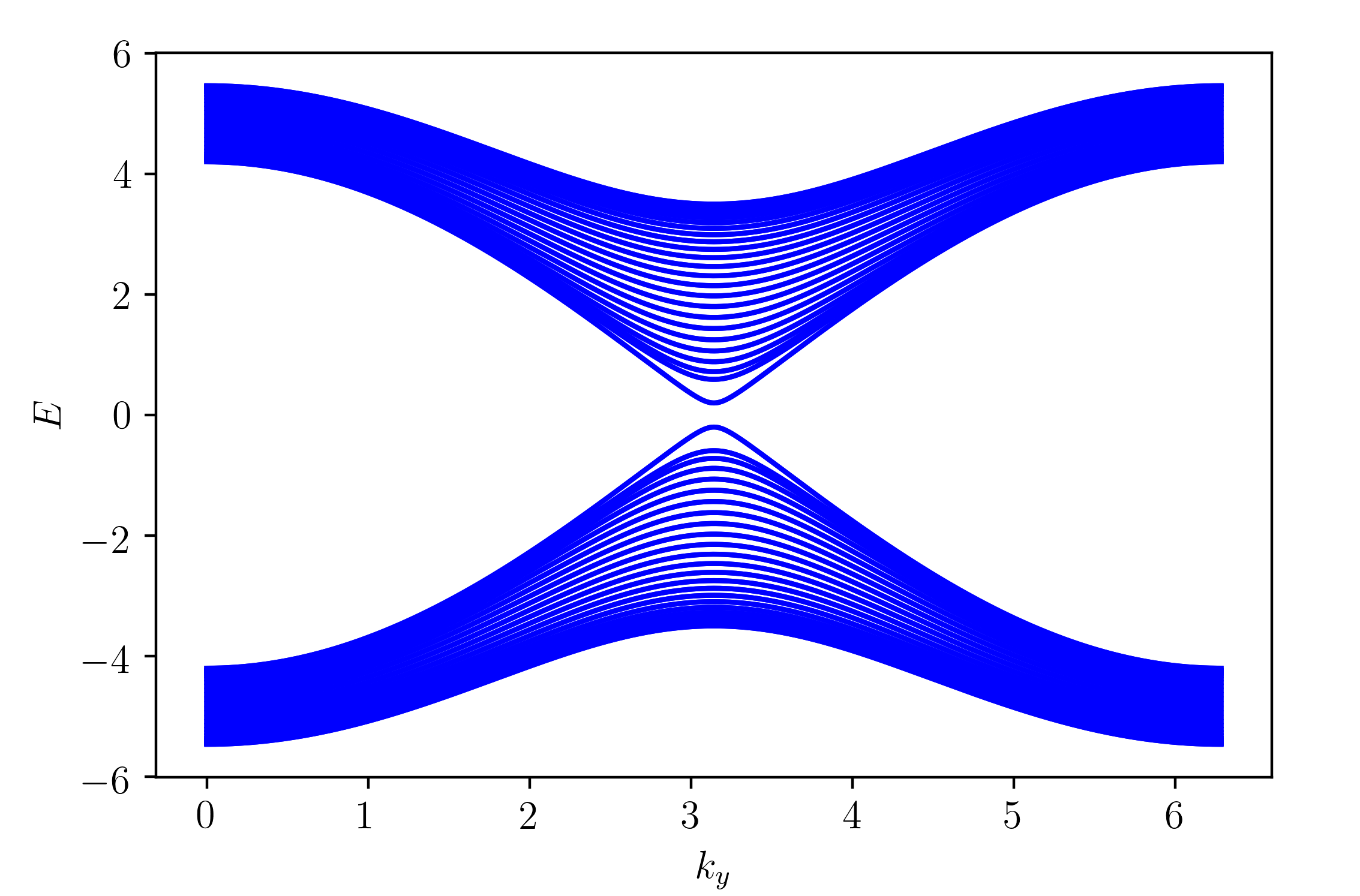}
         \caption{}
         \label{5.2}
     \end{subfigure}
\caption{{{(a) is the bandstructure for the semi-infinite configuration (Eq \ref{10}). Hamiltonian $H_x(k_x)$ is topological in this case. The values of the parameters are $|R_h|=1.5$, $|R_v|=2.0$. All the other parameters are unity. A gap closing transition is again observed at this point. (b) shows the opening of the bulk gap as $R_v$ exceeds a value of $2$. Here the value of $R_v=2.2$. All other parameters are kept the same.}}} 
 \label{fig5}
\end{figure}
For the two dimensional Hamiltonian (Eq. \ref{4}), we first look for a gap-closing transition in the bulk energy spectrum. The topological invariant for this phase is given by $\nu_{2D}=\nu_x\nu_y$, where $\nu_x$ and $\nu_y$ correspond to winding numbers pertaining to $H_x(k_x)$ and $H_y(k_y)$ respectively. The total winding number is non-zero only when both $H_x(k_x)$ and $H_y(k_y)$ are in the topological phase that is $\nu_x=\nu_y=1$.\par The real space Hamiltonian for the 2D Creutz ladder is,
\begin{align}
\begin{split}
\label{7}
    H &= \sum_{\vec{R}} iL_v(a_{\vec R}^{\dagger}a_{\vec R+\hat y}-d_{\vec R}^{\dagger}d_{\vec R+\hat y}+c_{\vec R}^{\dagger}c_{\vec R+\hat y}-b_{\vec R}^{\dagger}b_{\vec R+\hat y}) \\&+ R_v( a_{\vec R}^{\dagger}b_{\vec R} + c_{\vec R}^{\dagger}d_{\vec R}) + D_v(b_{\vec R}^{\dagger}a_{\vec R+\hat y} + b_{\vec R+\hat y}^{\dagger}a_{\vec R}\\&+c_{\vec R}^{\dagger}d_{\vec R+\hat y} + c_{\vec R+\hat y}^{\dagger}d_{\vec R}) + \sum_{\vec{R}}L_h(b_{\vec{R}}^{\dagger}a_{\vec{R}+\hat x}-b_{\vec{R}+\hat x}^{\dagger}a_{\vec{R}}\\&+c_{\vec{R}}^{\dagger}d_{\vec{R}+\hat x}-c_{\vec{R}+\hat x}^{\dagger}d_{\vec{R}})+iR_h(a_{\vec{R}}^{\dagger}d_{\vec R}+c_{\vec{R}}^{\dagger}b_{\vec R}) \\&+ iD_h(a_{\vec R+\hat x}^{\dagger}d_{\vec{R}}+a_{\vec{R}}^{\dagger}d_{\vec R+\hat x}+c_{\vec R+\hat x}^{\dagger}b_{\vec{R}}+c_{\vec{R}}^{\dagger}b_{\vec R+\hat x})+h.c.
\end{split}
\end{align}
Here $L_v$, $R_v$ and $D_v$ ($L_h$, $R_h$ and $D_h$) correspond to hopping amplitudes associated with the vertical (horizontal) Creutz ladder. It is to be noted that vertical Creutz ladder refers to $H_y(k_y)$. Similarly $H_x(k_x)$ is referred to as horizontal Creutz ladder. Interestingly now, the sublattice degree of freedom increases from being two in the original Creutz ladder to four in our two dimensional model. Here $a,b,c,d$ correspondingly denote the sublattice degrees of freedom for the two dimensional model. 
\begin{center}\section{\label{sec:level3}Results and discussions}\end{center}
We start by studying the phase diagram defined by the total winding number $\nu_{2D}$ (Fig. \ref{fig13}). It is evident that $\nu_{2D}$ is non-zero only in the central square-like region of the phase diagram where both $H_x(k_x)$ and $H_y(k_y)$ denote topological phases. We plot the energy eigenvalues corresponding to $R_h=R_v=1.5$ and $L_h=L_v=D_h=D_v=1$. The plot clearly shows the presence of the zero energy modes distinctly separated from the bulk (Fig. \ref{fig2}). The zero energy modes are four fold degenerate. These modes deviate from zero energy when either $H_x(k_x)$ or $H_y(k_y)$ or both progressively move towards the trivial phase.
A gap closing transition occurs in the bulk energy spectrum when $R_v=R_h=\pm 2$ (all other parameters are kept at unity). Our model shows the presence of two different topological phases. When the ladders along both the $x$ and $y$ directions are in the topological phase ($|R_h|,|R_v|<2$), we observe the presence of corner modes in the system under open boundary condition (OBC). We call this phase as $T_1$. This phase is characterized by a non-zero value of $\nu_{2D}$ (that is $\nu_{2D}=1$). The gap closing of the bulk energy spectrum at $R_v=R_h=\pm 2$ leads to a completely trivial phase where all the states are delocalized. Now, when only either of the $H_x(k_x)$ or $H_y(k_y)$ is in a topological phase and the other one is trivial, that is $\nu_x=1$ (or $\nu_y=1)$ and $\nu_y=0$ (or $\nu_x=0$), we find the existence of the edge modes. The energy of these edge modes is separated from the bulk. However despite being localized, they are not at zero energy. We call this the phase $T_2$. The edge modes are found along the edge perpendicular to the direction $i$ (where $i\in x,y$) for the Hamiltonian $H_i(k_i)$ which is in the topological phase. The phase $T_2$, is accordingly characterized by the winding number $\nu_i$. The bulk energy spectrum remains completely gapped during the transition from the first to the second topological phase ($T_1\rightarrow T_2$).\par In order to aid our understanding that there is no bulk gap closing transition in going from $T_1$ to $T_2$, we write down the square of the Hamiltonian, 
\begin{align}
\begin{split}
H^2(k_x,k_y)&=H_x(k_x)^2\otimes{\sigma_y^{y}}^2 + H_x(k_x)\mathbb{I}_x\otimes\sigma_y H_y(k_y)\\&+\mathbb{I}_xH_x(k_x)\otimes H_y(k_y)\sigma_y + \mathbb{I}_x\otimes H_y(k_y)^2\\&=H_x(k_x)^2\otimes{\sigma_y^{y}}^2+\mathbb{I}_x\otimes H_y(k_y)^2 \\&=H_x(k_x)^2\otimes\mathbb{I}_y+\mathbb{I}_x\otimes H_y(k_y)^2
\end{split}
\end{align}
Here, it is obvious that for $H(k_x,k_y)$ to have a gap closing, both $H_x(k_x)$ and $H_y(k_y)$ should take zero energy eigenvalues. Chiral symmetry ensures this property. However, it is to be noted that $\nu_{2D}$ changes its value during this transition ($T_1\rightarrow T_2$) inspite of no gap-closure occuring in the bulk energy spectrum. \par We, now consider a ribbon-like configuration of this model with periodic boundary condition (PBC) along the $x$ direction and the $y$ direction case-wise and study the behaviour of the band structures as a function of the parameters $R_h$ and $R_v$ respectively. 
\par
First we take a ribbon with periodic boundary condition (PBC) along the $x$ direction. The corresponding Hamiltonian is given as,
\begin{align}
\begin{split}
\label{9}
H(k_x)&= \mathbb{I}_{n_y}\otimes[H_x(k_x)\otimes\sigma^{y}_{y}] + \sum_{\vec R} \Big{[}iL_v(a_{\vec R}^{\dagger}a_{\vec R+\hat y}\\&-d_{\vec R}^{\dagger}d_{\vec R+\hat y}+c_{\vec R}^{\dagger}c_{\vec R+\hat y}-b_{\vec R}^{\dagger}b_{\vec R+\hat y})+ R_v( a_{\vec R}^{\dagger}b_{\vec R} \\&+ c_{\vec R}^{\dagger}d_{\vec R})+ D_v(b_{\vec R}^{\dagger}a_{\vec R+\hat y}+ b_{\vec R+\hat y}^{\dagger}a_{\vec R}+c_{\vec R}^{\dagger}d_{\vec R+\hat y} \\&+ c_{\vec R+\hat y}^{\dagger}d_{\vec R}) +h.c.\Big{]}
\end{split}
\end{align}
Here $\mathbb{I}_{n_y}$ denotes an identity matrix with $n_y$ being the number of lattice sites along the $y$ direction. For this configuration it is observed that the band structure undergoes a gap-closing transition at $|R_h|=2$ (It may be noted that $H_y(k_y)$ is kept in the topological phase all the while). 
Next, we consider a ribbon with PBC along the $y$ direction. The corresponding Hamiltonian is expressed as,
\begin{align}
\begin{split}
\label{10}
H(k_y)&=\mathbb{I}_{n_x}\otimes[\mathbb{I}_x \otimes H_y(k_y)] + \sum_{\vec{R}} \Big [L_h(b_{\vec R}^{\dagger}a_{\vec R+ \hat x}\\&-b_{\vec R + \hat x}^{\dagger}a_{\vec R}+c_{\vec R}^{\dagger}d_{\vec R+ \hat x}-c_{\vec R + \hat x}^{\dagger}d_{\vec R}) + iR_h(a_{\vec R}^{\dagger}d_{\vec R} \\&+c_{\vec R}^{\dagger}b_{\vec R})+ iD_h(a_{\vec R + \hat x}^\dagger d_{\vec R} + a_{\vec R}^\dagger d_{\vec R + \hat x}+c_{\vec R + \hat x}^\dagger b_{\vec R}  \\&+ c_{\vec R}^\dagger b_{\vec R + \hat x}) + h.c. \Big]
\end{split}
\end{align}
Again $\mathbb{I}_{n_x}$ denotes an identity matrix with $n_x$ being the number of lattice sites along the $x$ direction. For this configuration it is observed that the band structure undergoes a gap-closing transition at $|R_v|=2$ (As earlier, in this case $H_x(k_x)$ remains in the topological phase throughout). From this discussion it is implied that even though the bulk remains gapped during the $T_1\rightarrow T_2$ phase change, the ribbon like configuration shows the necessary transition. $\nu_{2D}$ goes from being topological ($\nu_{2D}=1$) to trivial ($\nu_{2D}=0$) implying that the corner states do not exist anymore. However this second topological state $T_2$ can be characterized, by $\nu_x$ or $\nu_y$ accordingly. $\nu_x$ (and similarly $\nu_y$) corresponds to the winding number that characterizes $H_x(k_x)$ (or $H_y(k_y)$). The prescription is the same as given in Eq. \ref{6}. It is to be noted that $H(k_x,k_y)$ has been constructed such that it preserves the chiral symmetry.
\begin{figure}
\begin{subfigure}[b]{\columnwidth}
         \centering
         \includegraphics[height=65mm,width=\columnwidth]{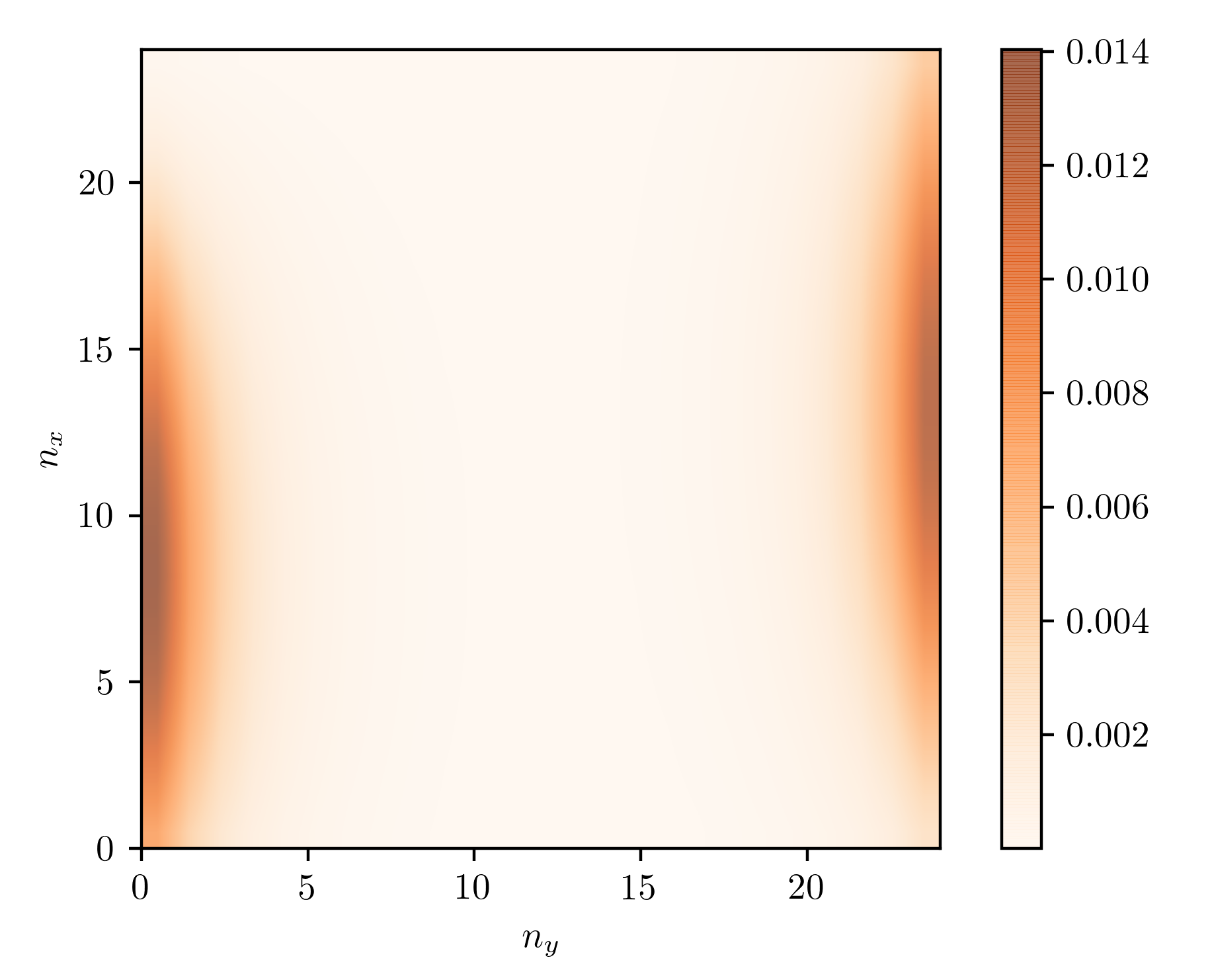}
         \caption{}
         \label{6.1}
     \end{subfigure}
\begin{subfigure}[b]{\columnwidth}
         \centering
         \includegraphics[height=65mm,width=\columnwidth]{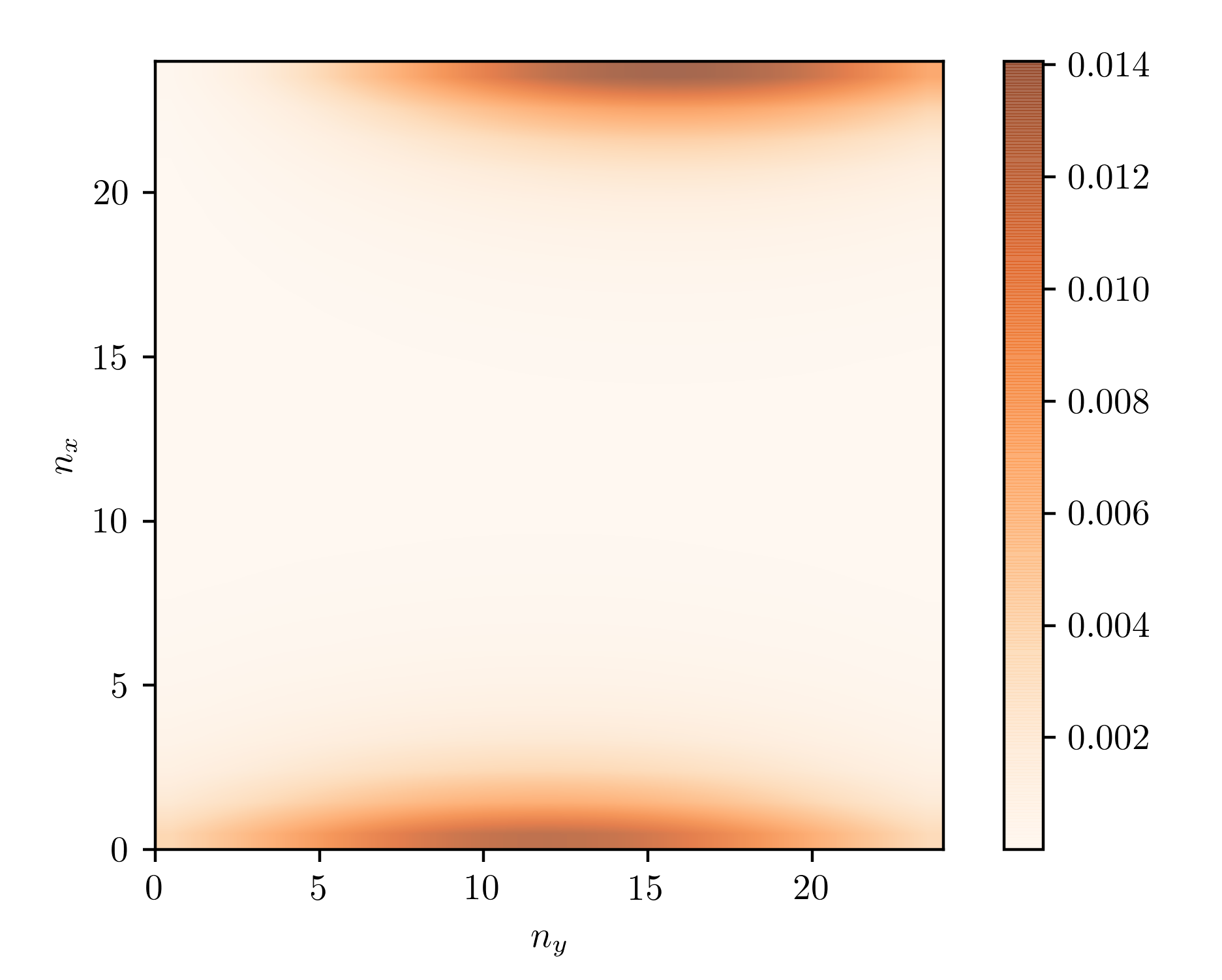}
         \caption{}
         \label{6.2}
     \end{subfigure}
\caption{{(a) and (b) refer to the edge modes for the topological phase $T_2$. In (a) $H_y(k_y)$ being topological, we find edge modes along the side perpendicular to the y direction. In this case $R_v=1.5$ and $R_h=2.2$. (b) depicts a case similar to (a) where $H_x(k_x)$ is in the topological phase. Here $R_h=1.5$ and $R_v=2.2$. Evidently edge modes are found along the y direction.}} 
 \label{fig6}
\end{figure}
The original one dimensional chiral Creutz ladder shows a band gap closing when $|R|=2$ (The values of the other parameters, namely $D$ and $L$, are maintained at a value of 1). Flat band spectrum is obtained for $R=0,L=1,D=1,\theta=\frac{\pi}{2}$. Similar characteristics are extrapolated for the 2D model as well and this is shown in Fig. \ref{fig1}. The Hamiltonian has a two fold degeneracy. A gap closing transition occurs when $|R_h|=|R_v|=2$ (Fig. \ref{1.1}). The band gap closes at several points in the Brillouin zone (BZ) namely, $(k_x,k_y)=(-\pi,-\pi),(\pi,-\pi),(-\pi,\pi),(\pi,\pi)$.
\begin{figure}
          \begin{subfigure}[b]{\columnwidth}
         \centering
         \includegraphics[height=65mm,width=\columnwidth]{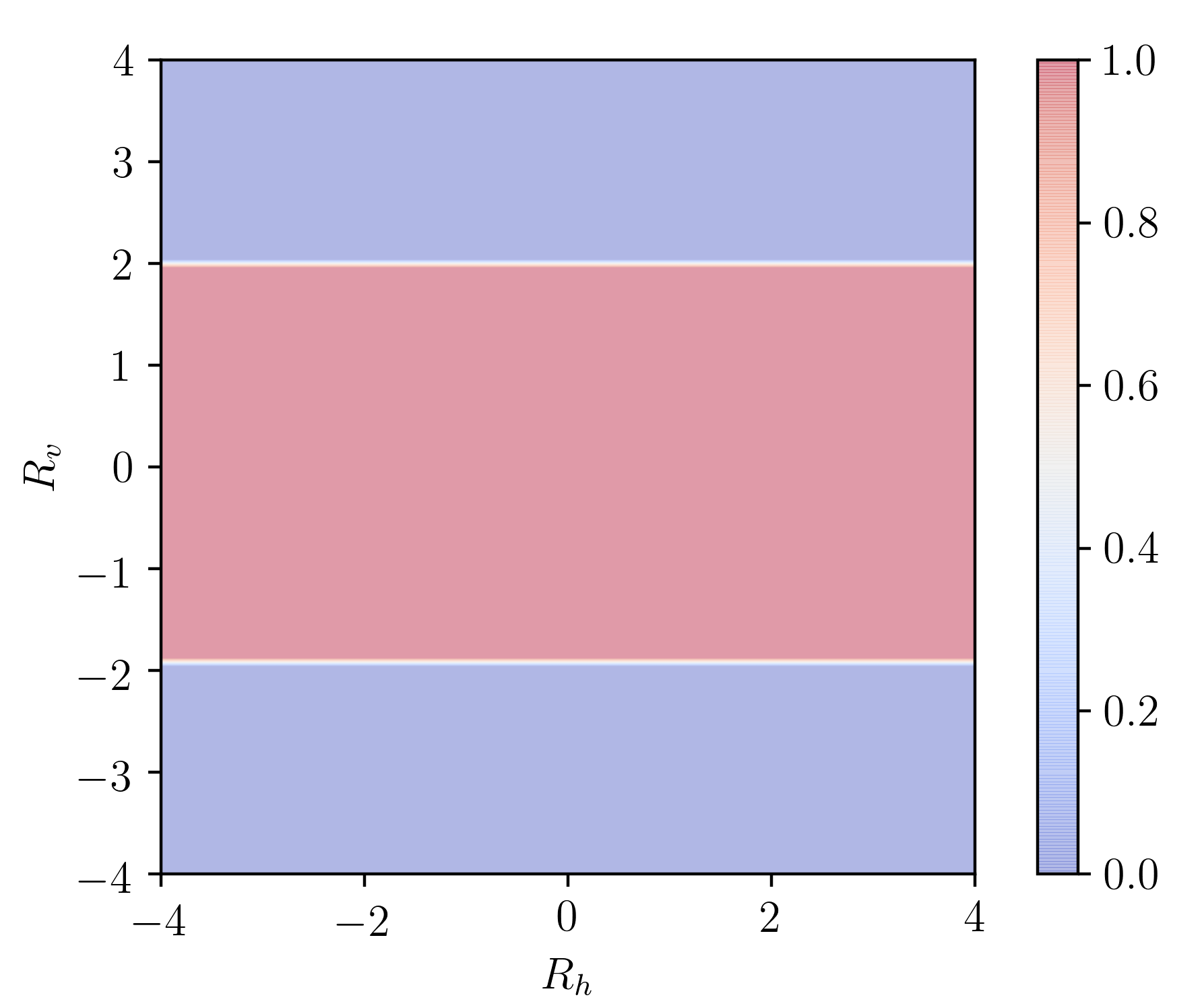}
         \caption{}
         \label{14.1}
     \end{subfigure}
     \begin{subfigure}[b]{\columnwidth}
         \includegraphics[height=65mm,width=\columnwidth]{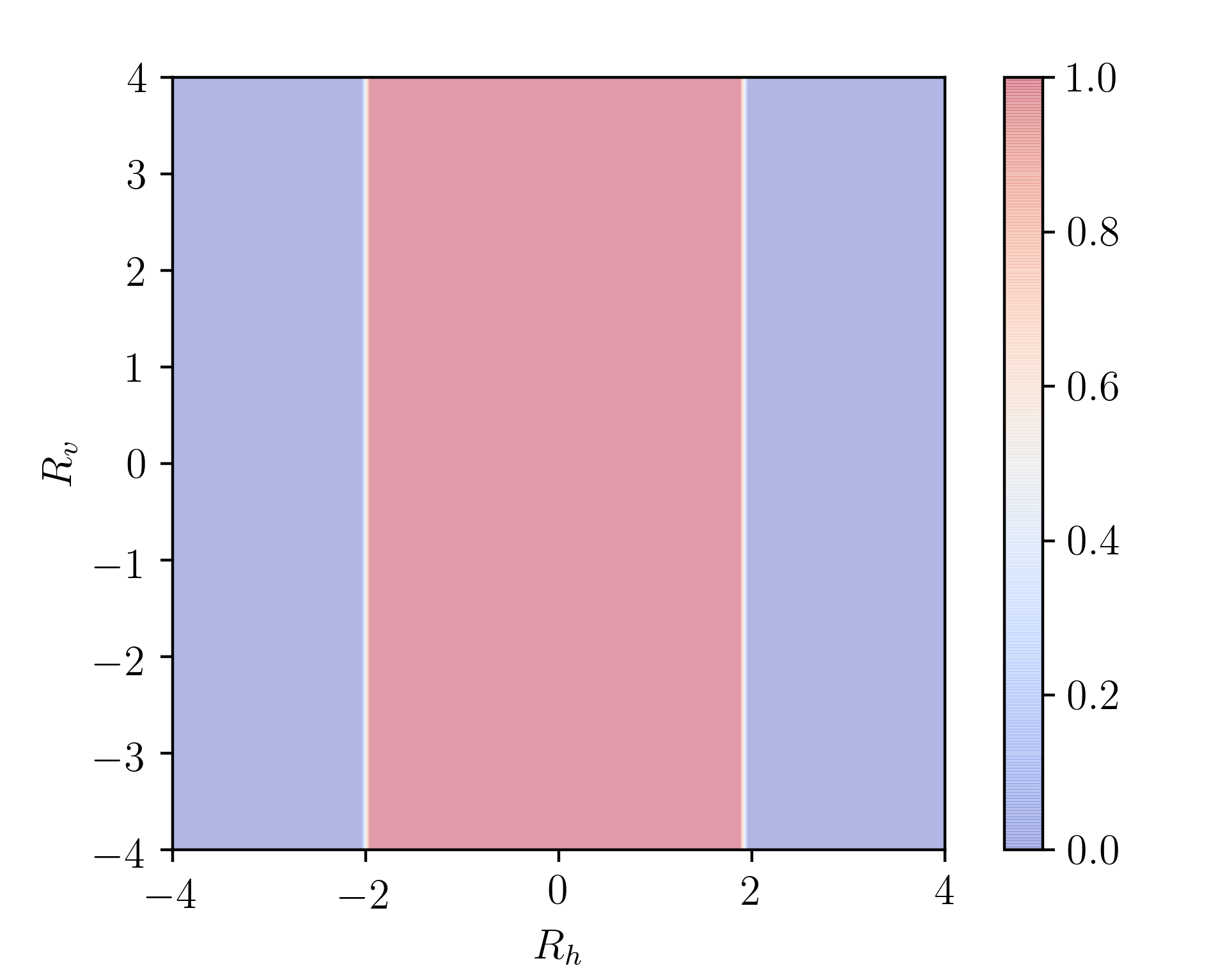}
         \caption{}
         \label{14.2}
     \end{subfigure}
 \caption{{(a) refers to the phase space plot for the topological invariant $\nu_y$ plotted with respect to $R_h$ and $R_v$. (b) depicts the phase space plot for $\nu_x$.}} 
 \label{fig14}
\end{figure}
Moreover, a flat band (same as the 1D model) is seen when $R_h=R_v=0$ (Fig. \ref{1.2}).\par To visualise the corner modes corresponding to the topological phase $T_1$, we first take a finite lattice in the form of a square with $25\times25$ unit cells along the $x$ and the $y$ directions. The corner states are mostly confined at the parameter values corresponding to the flat band dispersion (Fig \ref{fig3}). This is due to Aharanov Bohm localization (ABL) effect. When $R_h$ and $R_v$ vanish, the particles have fewer paths to hop to its neighbours. Moreover for $\theta=\frac{\pi}{2}$ the particles interfere destructively causing them to be localized at their own plaquettes. The localization in the Creutz ladder is a unique effect of quantum interference and topological protection. Once the zero energy states are created at the flat band point, the chiral symmetry forces them to remain there even when the parameters are changed. For these states to merge into the bulk, a gap closing transition is required.\par Now, we focus on the second topological phase, namely $T_2$. The bulk spectrum does not close during the transition ($T_1 \rightarrow T_2$) giving a false implication that there is no transition taking place there. Table \ref{T1} provides information about the parametric regime corresponding to the different phases of the model. It also elaborates on the topological invariants associated with these phases.
\begin{table}[h!]
\begin{center}
\setlength{\tabcolsep}{10pt} 
\renewcommand{\arraystretch}{1.5}
\begin{tabular}{|| c | c | c | c | c | c ||}
\hline
 &$\nu_x$&$\nu_y$&$\nu_{2D}$&$|R_h|$&$|R_v|$ \\
\hline
\hline
$T_2$&$0$&$1$&$0$&$>2$&$<2$\\
\hline
$T_2$&$1$&$0$&$0$&$<2$&$>2$\\
\hline
$T_1$&$1$&$1$&$1$&$<2$&$<2$\\
\hline
\end{tabular}
\caption{{Table depicting various topological phases of the two dimensional Creutz ladder. $\nu_{2D}$, $\nu_x$, $\nu_y$ correspond to the topological invariants.}}
\label{T1}
\end{center}
\end{table}
Evidently, $\nu_{2D}$ is inadequate to characterize the topological phase $T_2$. As already mentioned, we investigate two different semi-infinite configurations with OBC first along the $y$ direction (when $H_y(k_y)$ is topological) and then along the $x$ direction (when $H_x(k_x)$ is topological). The bandstructure is accordingly plotted as a function of $k_x$ or $k_y$ (Fig. \ref{fig4} and Fig. \ref{fig5} respectively). The gap closes at $|R_h|=2$ and $|R_v|=1.5$ in Fig \ref{4.1}, while in Fig. \ref{5.1}, the same occurs for $|R_h|=1.5$ and $|R_v|=2$. We take $25$ unit cells along the width of the ribbon. This spectrum shows gap closure as a function of $R_h$ as shown in Fig. \ref{fig4} (or $R_v$ as shown in Fig. \ref{fig5}) as the system goes from one topological phase to another, that is, $T_1\rightarrow T_2$.\par
Now, we look at the invariants characterizing the two different $T_2$ phases. First, we study the case where $|R_v|<2,|R_h|\ge2$. For a finite square lattice with a size as above, owing to the topological nature of $H_y(k_y)$, the phase is characterized by a non-zero value of $\nu_y$. The edges perpendicular to the $y$ direction show the presence of edge modes as long as $|R_v|<2$ (Fig. \ref{6.1}). The states merge into the bulk as $|R_v|\ge2$. For the second case, $|R_h|<2,|R_v|\ge2$. Here the boundary perpendicular to the $x$ direction shows the presence of edge modes as long as $|R_h|<2$ (Fig. \ref{6.2}). The topological non-triviality vanishes when $|R_h|\ge2$. The topological invariant characterizing this phase is $\nu_x$. We plot three different phase space plots corresponding to the two topological phases $T_1$ and $T_2$ (Fig. \ref{fig13} and Fig. \ref{fig14}). We study the variation of the three different winding numbers ($\nu_{2D},\nu_x,\nu_y$) with respect to the variation of the parameters $R_h$ and $R_v$. They depict the region where the invariants are non-trivial and corner or edge modes are expected to exist. The presence and the absence of the edge or the corner modes is found to be in agreement with the value of the topological invariants.
\begin{center}{\section{\label{sec:level4}Conclusion}}\end{center}
The Creutz ladder in its original version,  has four degrees of freedom namely, $R, L, D$ as the hopping amplitudes and $\theta$ which denotes magnetic flux. A combination of these control the topological properties of the model. Similar behaviour of localization protected by AB caging and topology is observed when the Creutz ladder is extrapolated to two dimensions while maintaining the chiral symmetry. Two different configurations, one being a torus (PBC along both the $x$ and the $y$ directions) and the other being a ribbon (PBC along either $x$ or $y$ direction) is studied. A bulk gap closing transition takes us directly from the higher order topological phase ($T_1$), that exhibits corner modes, to a trivial phase where all the states are delocalized. However, a gap closure in the ribbon-like configuration takes us from the higher order phase ($T_1$) to a first order phase ($T_2$) where the edge states can be seen. A new topologial invariant $\nu_{2D}$ is defined to quantify the corner modes in the $T_1$ phase. The non-triviality of the $T_2$ phase is characterized by the winding number of the Hamiltonian $H_i(k_i)$, where $i\in x$ or $y$, which is topological. The results on Creutz ladder would be promising in the field of Quantum Information because of the presence of robust edge or corner states, which can be tuned efficiently by changing its parameters.
\bibliographystyle{ieeetr}
\bibliography{ref.bib}
\end{document}